
\input phyzzx

\def\msbar{\overline{\rm MS}}
\def\lamqcd{\Lambda_{\rm QCD}}
\def\lambar{\ol\Lambda}
\def\leff{{\cal L}_{\rm eff}}
\def\lfull{{\cal L}_{\rm full}}
\def\lkin{{\cal L}_{\rm kin}}
\def\llight{{\cal L}_{\rm light}}
\def\dm{\delta m}
\def\ol{\overline}
\def\im{{\rm i}}
\def\e{{\rm e}}
\def\b{{\rm b}}
\def\c{{\rm c}}
\def\q{{\rm q}}
\def\fr#1#2{\textstyle{#1\over#2}}
\def\CO{{\cal O}}
\def\eps{\varepsilon}
\def\epslash{\not{\hbox{\kern-2.3pt $\eps$}}}
\def\kslash{\not{\hbox{\kern-2.3pt $k$}}}
\def\vslash{\not{\hbox{\kern-2.3pt $v$}}}
\def\Dslash{\not{\hbox{\kern-4pt $D$}}}
\def\vcvb{v_\c\cdot v_\b}
\def\braa#1{\langle\, #1 \,}
\def\kett#1{\, #1\, \rangle}
\def\mudmu#1{\mu{\partial{#1}\over\partial\mu}}
\def\to{\rightarrow}

\Pubnum={SLAC--PUB--5771\cr UCSD/PTH 92-09}
\date{March 1992}
\pubtype{T/E}
\titlepage
\title{The Residual Mass Term in the Heavy Quark Effective
Theory\footnote\star{Work supported by the Department of Energy,
contracts DE--AC03--76SF00515 and DE--FG03--90ER40546.}}
\author{Adam F. Falk and Matthias Neubert}
\SLAC
\andauthor{Michael Luke}
\address{Department of Physics B-019\break University of
California at San
Diego\break 9500 Gilman Drive, La Jolla, CA 92093-0319}
\abstract
\singlespace

We reformulate the heavy quark effective theory in the presence of a
residual mass term, which has been taken to vanish in previous analyses.
While such a convention is permitted, the inclusion of a residual mass
allows us to resolve a potential ambiguity in the choice of the expansion
parameter which defines the effective theory.
We show to subleading order
in the mass expansion that physical quantities computed in the effective
theory do not depend on the expansion parameter.
\submit{Nuclear Physics B}
\endpage


\REF\lepa{W.E. Caswell and G.P. Lepage, {\sl Phys.\ Lett.} {\bf B167}
(1986) 437;\nextline
G.P. Lepage and B.A. Thacker, {\sl Nucl.\ Phys.} {\bf B} (Proc.\ Suppl.)
{\bf 4} (1988) 199.}

\REF\volo{M.B. Voloshin and M.A. Shifman, {\sl Yad.\ Fiz.} {\bf 45}
(1987) 463 [{\sl Sov.\ J.\ Nucl.\ Phys.} {\bf 45} (1987) 292],
{\sl Yad.\ Fiz.} {\bf 47} (1988) 511 [{\sl Sov.\ J.\ Nucl.\ Phys.}
{\bf 47} (1988) 511.]}

\REF\poli{H.D. Politzer and M.B. Wise, {\sl Phys.\ Lett.} {\bf B206}
(1988) 681, {\sl Phys.\ Lett.} {\bf B208} (1988) 544.}

\REF\isgu{N. Isgur and M.B. Wise, {\sl Phys.\ Lett.} {\bf B232} (1989)
113, {\sl Phys.\ Lett.} {\bf B237} (1990) 527.}

\REF\eich{E. Eichten and B. Hill, {\sl Phys.\ Lett.} {\bf 234} (1990)
511, {\sl Phys.\ Lett.} {\bf 243} (1990) 427.}

\REF\huss{F. Hussain \etal, {\sl Phys.\ Lett} {\bf B249} (1990) 295;
\nextline
F. Hussain, J.G. K\"orner, M. Kr\"amer and G. Thompson, {\sl Z.\ Phys.}
{\bf C51} (1991) 321.}

\REF\mann{T. Mannel, W. Roberts and Z. Ryzak, {\sl Phys.\ Lett.}
{\bf B255} (1991) 593, {\sl Nucl.\ Phys.} {\bf B355} (1991) 38.}

\REF\neub{M. Neubert, {\sl Phys.\ Lett.} {\bf B264} (1991) 455.}

\REF\georgi{H. Georgi, {\sl Phys.\ Lett.} {\bf B240} (1990) 447.}

\REF\korn{J.G. K\"orner and G. Thompson, {\sl Phys.\ Lett.} {\bf B264}
(1991) 185.}

\REF\MRR{T. Mannel, W. Roberts and Z. Ryzak, Harvard preprint
HUTP--91/A017 (1991).}

\REF\fggw{A.F. Falk, H. Georgi, B. Grinstein and M.B. Wise, {\sl Nucl.\
Phys.} {\bf B343} (1990) 1.}

\REF\mike{M.E. Luke, {\sl Phys.\ Lett.} {\bf B252} (1990) 447.}

\REF\fgl{A.F. Falk, B. Grinstein and M.E. Luke, {\sl Nucl.\ Phys.}
{\bf B357} (1991) 185.}

\REF\models{N. Isgur and M.B. Wise, {\sl Phys.\ Rev.} {\bf D43} (1991)
819.}

\REF\Volker{M. Neubert and V. Rieckert,
Heidelberg preprint HD--THEP--91--6
(1991).}

\REF\buch{M. Neubert, V. Rieckert, B. Stech and Q.P. Xu, Heidelberg
preprint HD--THEP--91--28 (1991), to appear in {\sl ``Heavy Flavours''},
edited by A.J. Buras and M. Lindner, Advanced Series on Directions in
High Energy Physics, World Scientific Publishing Co.}

\REF\rady{A.V. Radyushkin, {\sl Phys.\ Lett.} {\bf B271} (1991) 218.}

\REF\SR{M. Neubert, SLAC preprint SLAC--PUB--5712 (1991), to appear in
{\sl Phys.\ Rev.} {\bf D45} (1992).}

\REF\matthias{M. Neubert, SLAC preprint SLAC--PUB--5770 (1992).}

\REF\BBBD{E. Bagan, P. Ball, V.M. Braun and
H.G. Dosch, Heidelberg preprint
HD--THEP--91--36 (1991).}

\REF\allt{C. R. Allton \etal, {\sl Nucl.\ Phys.} {\bf B349} (1991)
598.}

\REF\alex{C. Alexandrou \etal, {\sl Phys.\ Lett.} {\bf B256} (1991)
60.}

\REF\bern{C. Bernard, A.X. El-Khadra, A. Soni,
{\sl Phys.\ Rev.} {\bf D43}
(1991) 2140.}

\REF\maia{L. Maiani, {\sl Helv.\ Phys.\ Acta} {\bf 64} (1991) 853.}

\REF\mms{L. Maiani, G. Martinelli and C.T. Sachrajda,
Southampton preprint
SHEP 90/91--32 (1991);\nextline
A. Morelli, Brookhaven preprint BNL--47091 (1992).}

\REF\FG{A.F. Falk and B. Grinstein, {\sl Phys.\ Lett.} {\bf B247}
(1990) 406.}

\REF\pert{A.F. Falk and B. Grinstein, {\sl Phys.\ Lett.}
{\bf B249} (1990)
314.}

\REF\jimu{X. Ji and M.J. Musolf, {\sl Phys.\ Lett.} {\bf B257} (1991)
409.}

\REF\broa{D.L. Broadhurst and A.G. Grozin, {\sl Phys.\ Lett.} {\bf B267}
(1991) 105.}

\REF\korc{G.P. Korchemsky and A.V. Radyushkin,
{\sl Nucl.\ Phys.} {\bf B283}
(1987) 342, Marseille preprint CPT--91/P.2629 (1991).}

\REF\QCD{M. Neubert, Heidelberg preprints HD--THEP--91--4 (1991),
to appear
in {\sl Nucl.\ Phys.} {\bf B}, and HD--THEP--91--30 (1991).}

\REF\bj{J.D. Bjorken, talk given at Les Rencontres de la Valle d'Aoste La
Thuile, Aosta Valley, Italy, March 1990, SLAC preprint SLAC--PUB--5278
(1990).}

\REF\adam{A.F. Falk, SLAC preprint SLAC--PUB--5689 (1991), to appear
in {\sl Nucl.\ Phys.} {\bf B}.}

\singlespace
\parskip=0pt

\section{1. Introduction}

There has been much recent interest in the limit of QCD in which the
mass of a heavy quark is taken to be much larger than the characteristic
scale $\lamqcd$ of the strong interactions [\lepa-\fgl].
In this limit it is natural to consider an expansion of
the QCD lagrangian in inverse powers of the heavy quark mass, as well
as kinematics in which the heavy quark, even though bound into a
hadron, is almost on shell.  Such an expansion has been used
to lowest order in models of heavy hadrons [\models,\Volker], in
QCD sum rule calculations [\buch-\BBBD], and in lattice gauge
theory [\allt-\maia]. However, to
put the predictions of this heavy quark effective theory (HQET) on firm
footing it will be necessary to perform computations to subleading
order in the expansion in inverse mass $1/m_Q$.  But here arises a
potential ambiguity, namely in what
mass should one expand:  the pole mass
computed in perturbation theory, the $\msbar$ mass, the mass of the
physical hadron, or some other parameter?
Our intuition tells us that nothing physical can
depend on this choice, and this is in fact the case.  It is the purpose
of this note to demonstrate in detail how physical matrix elements are
insensitive to variations in the expansion parameter.

Throughout this paper we will use dimensional regularization to cut
off the ultraviolet divergences which arise in the HQET.  We will not
address the important issue of whether power divergences, which
appear when a dimensionful regulator is employed, require additional
nonperturbative subtractions in the effective theory [\mms].
However, the arguments given here are sufficient to show that if
such effects are important and introduce additional uncertainties into
computations, they are unrelated to
ambiguities in the choice of the expansion parameter which defines the
effective theory.  We feel that it is important to disentangle these
two issues.

The potential ambiguity in the choice of expansion parameter arises
because in a confining theory such as QCD there is no true ``pole'' mass
which can be assigned to a heavy quark.
In the absence of such a canonical
choice, a variety of perturbative prescriptions (\eg, the pole mass to a
given order, or the mass renormalized at some scale) compete with
``physical'' prescriptions such as the mass of the lightest hadron
containing the heavy quark, or this mass minus some fixed number.

In this paper we will show that these various choices correspond in the
effective theory to various values of a new parameter, the ``residual
mass'' $\dm$. Section~2 is devoted to the construction of this more
general form of the HQET. In Section~3 we discuss, to subleading order
in $1/m_Q$, hadronic matrix elements of currents containing two heavy
quarks. Currents with one heavy and one light quark are addressed in
Section~4. Section~5 contains a brief summary of our results.

\section{2. HQET with a Residual Mass Term}

Let us fix any reasonable choice for the heavy quark mass and denote
it by $m_Q$. Each value of
$m_Q$ yields a different effective theory; it is the parameter which
defines, once and for all, the relationship between the HQET and full
QCD. Following refs.~[\georgi-\MRR], we derive the effective theory as
an expansion in $1/m_Q$. We rewrite the heavy quark momentum $P_\mu$ as
$$ P_\mu=m_Qv_\mu+k_\mu ~,
   \eqn\modef$$
where $v_\mu$ is the four-velocity of the hadron containing the heavy
quark; in the $m_Q\to\infty$ limit it is also the velocity of the heavy
quark itself.  Note that $k_\mu$ does not scale with $m_Q$; in fact, it
is this property of separating out the
part of $P_\mu$ that scales with the heavy quark mass which
characterizes a ``reasonable'' choice of $m_Q$.\footnote{*}{Note
that the $\msbar$ mass $m_Q(m_Q)$ is not a ``reasonable'' choice, since
in this case $k_\mu$ contains a piece proportional to
$\alpha_s(m_Q)m_Q v_\mu$.}   We also
define effective fields $h_Q(x)$ in
terms of the original fields $Q(x)$ by
$$ h_Q(x)=\exp(\im m_Q\vslash\, v\cdot x )\,Q(x) ~.
   \eqn\fielddef$$
These fields create and annihilate heavy quarks and
antiquarks moving at velocity $v_\mu$. We further specialize to heavy
quarks by imposing the condition $\vslash h_Q=h_Q$.  To lowest order in
the $1/m_Q$ expansion, the
effective lagrangian is obtained by matching matrix elements in QCD
with those in the effective theory.  We find
$$ \leff=\ol h_Q(\im v\cdot D)h_Q-\dm\,\ol h_Q h_Q
   =\ol h_Qv^\mu(\im D_\mu-\dm v_\mu) h_Q ~,
   \eqn\lefflowest$$
where $D_\mu=\partial_\mu-\im gA_\mu$ is the gauge-covariant derivative.

The term proportional to $\dm$ corresponds to a residual mass for
the heavy quark in the effective theory, of a size of order $\lamqcd$.
In general, such a term arises due to an incomplete cancelation of the
full theory mass by the field redefinition \fielddef. Changes in $m_Q$
will induce changes in this uncompensated mass:  if $m_Q\to
m_Q+\alpha$, then $\dm\to\dm-\alpha$. Similarly the residual momentum
changes according to $k_\mu\to k_\mu-\alpha v_\mu$. We may write these
relations in differential form as
$$ {\partial\dm\over\partial m_Q}=-1 ~,~~
   {\partial k_\mu\over\partial m_Q}=-v_\mu ~.
   \eqn\difrel$$
The residual mass term has been neglected in all previous derivations
based on the HQET. This is because $m_Q$
always has been fixed implicitly,
by imposing {\it post facto\/} the
condition $\dm=0$. Clearly there is some
choice of $m_Q$ which accomplishes this,
even if we cannot compute it.  In
other words, these derivations have fixed the equations of motion
of the heavy quark in the HQET to be $(\im v\cdot D)h_Q=0$, rather
than the more general condition $(\im v\cdot D-\dm)h_Q=0$.  Of course
setting $\dm=0$ is as valid a choice of $m_Q$ as any other; in fact, for
perturbative calculations it is by far the most convenient one. Yet
nowhere it is forced upon us.  Finally we note that although it is
possible to formulate the HQET with any mass $\dm$ of the order of
$\lamqcd$, if there is more than one heavy quark in the theory it is
convenient to choose $\dm$ to be the same for all of them, so as not to
violate the heavy quark symmetry explicitly at leading order.

Essentially the only modification which we have introduced into the
effective theory is the change in the equation of motion for the heavy
quark, which we may write in momentum space as
$$ v^\mu(k_\mu-\dm v_\mu + g A_\mu)\,h_Q=0 ~.
   \eqn\eqnmotion$$
This equation is used in the matching conditions which
determine the effective lagrangian to subleading order in $1/m_Q$, as
well as in the matching of currents in the full
theory onto currents in the effective theory.  Recall that we may use
the equations of motion here because the matching conditions are
obtained by requiring the equality of
{\it physical\/} matrix elements in the
two theories.

We have recalculated, to leading logarithmic order in perturbation
theory and to order $1/m_Q$, the operators and currents which arise in
the HQET with two heavy quarks, which we shall call for concreteness b
and c.  This corresponds to QCD plus a contact term which
describes the semileptonic decay of b to c quarks:
$$ \lfull=\overline\b(\im\Dslash-m_\b)\b +
   \overline\c(\im\Dslash-m_\c)\c+\ol\c\Gamma^\mu\b L_\mu+\llight ~,
   \eqn\lfull$$
where $\llight$ describes the interactions
of the light quarks and gluons.
Here $L_\mu$ is the charged lepton current (which includes constants such
as $G_F$ and $V_{\c\b}$), and we have written the left-handed heavy quark
current in the general form $\ol\c\Gamma^\mu\b$.  In the HQET, $\llight$
is unchanged, while the kinetic and interaction terms are expanded
in powers of $1/m_Q$. The strong interaction lagrangian for the heavy
quarks becomes
$$ \eqalign{
   \lkin &= \ol{h_\b}(\im v_\b\cdot D-\dm)h_\b+{1\over 2m_\b}
   \ol h_\b(\im D-\dm v_\b)^2 h_\b \cr
   &+ a_\b(\mu){g\over4m_\b}\ol h_\b\sigma_{\mu\nu} G^{\mu\nu}h_\b
   \;+\; (\b\to\c) ~, \cr}
   \eqn\lkineff$$
where we have ignored an operator whose matrix elements vanish by the
equations of motion. Such operators do not contribute to physical matrix
elements.  In leading logarithmic approximation,
$a_Q(\mu)=[\alpha_s(m_Q)/\alpha_s(\mu)]^{9/\beta}$,
where $\beta=33-2N_f$, and $N_f$ is the number of light quark flavors.
The $\mu$-dependence of the coefficients is required to
cancel the subtraction-point dependence of the renormalized operators
in the usual way.  We emphasize that $m_\b$ and $m_\c$ are {\it fixed\/}
($\mu$-independent) expansion parameters which define the effective
theory.  Finally, the effective lagrangian is written in terms of
renormalized heavy quark fields, which are related to the bare fields
by a factor $Z_Q^{1/2}=[\alpha_s(m_Q)/\alpha_s(\mu)]^{4/\beta}$.

The interaction term in the effective theory is fixed by the matching
of the currents $J=\ol\c\Gamma\b$ in the full theory onto linear
combinations of effective currents. Decomposing to order $1/m_Q$ the full
theory on-shell spinor $u$ in terms of the two-component spinor $u_h$
which describes the heavy quark in the effective theory,
$$ u=\Bigl\lbrack 1+{1\over2m_Q}(\kslash-\dm)
   \Bigr\rbrack\,u_h ~, \eqn\decompose$$
we obtain the matching conditions at tree level. Additional operators are
then induced by the renormalization group.  To leading logarithmic
order, the result is
$$ J_{\rm eff}=c_0(\mu)Q_0+c_1(\mu)Q_1+c_2(\mu)Q_2+c_3(\mu)Q_3
   +c_4(\mu)Q_4 ~,
   \eqn\jeff$$
where
$$ \eqalign{
   Q_0 &=\e^{\im\phi}\,\ol h_\c\Gamma h_\b ~, \cr
   Q_1 &=\e^{\im\phi}\,{1\over 2m_\c}\ol
   h_\c(-\im\overleftarrow\Dslash-\dm)
    \Gamma h_\b ~, \cr
   Q_2 &=\e^{\im\phi}\,{1\over 2m_\c}\ol
   h_\c(-\im v_\b\cdot\overleftarrow D-
    \dm\vcvb)\Gamma h_\b ~, \cr
   Q_3 &=\e^{\im\phi}\,{1\over 2m_\b}\ol
   h_\c\Gamma(\im\Dslash-\dm) h_\b ~,
    \cr
   Q_4 &=\e^{\im\phi}\,{1\over 2m_\b}\ol
   h_\c\Gamma(\im v_\c\cdot D-\dm \vcvb)
    h_\b ~. \cr}
   \eqn\operatordefs$$
We have again neglected operators whose matrix elements vanish by the
equations of motion. The phase $\phi=-(m_\b v_\b-m_\c v_\c)\cdot x$
compensates for the field redefinition \fielddef.
At tree level, $c_0=c_1=c_3=1$, $c_2=c_4=0$, while summing the leading
logarithms yields
$$ \eqalign{
   c_0(\mu) &=c_1(\mu)=c_3(\mu)=\left({\alpha_s(\widetilde m)\over
   \alpha_s(\mu)}\right)^{a_L} ~,\cr
   c_2(\mu) &=c_4(\mu)=-{16\over\beta}\left({r(w)-w\over w^2-1}\right)
   \left({\alpha_s(\widetilde m)\over
   \alpha_s(\mu)}\right)^{a_L}\ln\left(
   {\alpha_s(\widetilde m)\over\alpha_s(\mu)}\right) ~,\cr}
   \eqn\bdefs$$
where
$$ \eqalign{
   w &=\vcvb,\qquad a_L={8\over\beta} \big[ wr(w)-1 \big] ~,\cr
   r(w) &={1\over\sqrt{w^2-1}}\ln\big(w+\sqrt{w^2-1}\big) ~.\cr}
   \eqn\fundefs$$
These expressions arise when the transition from QCD to the
effective theory is implemented at a single scale $\widetilde m$, which in
this case is some mass between $m_\c$ and $m_\b$.  Improvements such
as including the full one loop matching
conditions, summing the logarithms
between $m_\c$ and $m_\b$ to resolve the ambiguity in $\widetilde m$, or
working to higher order in perturbation theory, are discussed in the
literature [\FG-\QCD].

 From \lkineff\ and \operatordefs\ it is apparent that in the effective
theory with a nonvanishing $\dm$ the covariant derivative acting on a
heavy quark field $h_Q(v)$ always appears in the combination
$$ \im{\cal{D}}_\mu(v) \equiv \im D_\mu - \dm v_\mu ~,~~
   \im\overleftarrow{\cal{D}}_\mu(v) \equiv
   \im\overleftarrow{D}_\mu + \dm v_\mu ~.
   \eqn\deriv $$
Note that the ``magnetic interaction'' operator in \lkineff\ is also
of this type, since $G_{\mu\nu}$ is proportional to $[{\cal{D}}_\mu,
{\cal{D}}_\nu]$. From \difrel\ we may conjecture that matrix elements
of the operator ${\cal{D}}$ do not depend on the choice of $m_Q$. We
shall now demonstrate that this is indeed the case.

\section{3. Hadronic Matrix Elements}

Let us evaluate the hadronic matrix elements of the weak
current in the effective theory with a residual mass term, repeating the
arguments of ref.~[\mike] in this more general context.  For simplicity,
we shall restrict ourselves to the
discussion of $1/m_\c$ corrections only.
The inclusion of terms proportional
to $1/m_\b$ is straightforward [\Volker].
We will also specialize, for purpose of illustration, to transitions
between $B^{(*)}$ and $D^{(*)}$ mesons. At next-to-leading order in the
heavy quark expansion, one has to include the $1/m_\c$ corrections to the
current and to the hadronic wave functions. Matrix elements of the
effective current \jeff\ can be evaluated
most concisely by using a covariant
trace formalism [\fggw,\mike,\bj,\adam],
$$ \eqalign{
  \braa{D^{(*)}(v_\c)}|\,\ol h_\c\Gamma h_\b\,|\kett{B^{(*)}(v_\b)}
  &= -\xi_0(w,\mu)\Tr\left[\,\ol M_D\Gamma M_B\,\right] ~,\cr
  \braa{D^{(*)}(v_\c)}|\,\ol h_\c
  \big[-\im\overleftarrow{\cal{D}}_\mu(v_\c)\big]
  \Gamma h_\b\,|\kett{B^{(*)}(v_\b)}
  &=-\Tr\left[\,\xi_\mu(v_\c,v_\b,\mu)\,\ol M_D\Gamma M_B\,\right] ~,\cr}
  \eqn\traceform$$
where pseudoscalar and vector meson states
are represented respectively as
$$ M(v)=-\sqrt{m_M}\,{1+\vslash\over2}\gamma^5 ~,\qquad
   M^*(v,\eps)=\sqrt{m_{M^*}}\,{1+\vslash\over2}\epslash ~.
   \eqn\statereps$$
The universal Isgur-Wise function $\xi_0$ is the only form factor that
appears in leading order of the $1/m_Q$ expansion [\isgu]. At
subleading order, there are additional functions. The most general
Lorentz-invariant decomposition of $\xi_\mu$ is
$$ \xi^\mu(v_\c,v_\b,\mu) = \xi_+(w,\mu)\,(v_\c^\mu+v_\b^\mu)
   + \xi_-(w,\mu)\,(v_\c^\mu-v_\b^\mu) - \xi_3(w,\mu)\,\gamma^\mu ~.
   \eqn\xidecom$$
The form factors $\xi_i$, which parametrize matrix elements in the
effective theory, are independent of the heavy quark masses. They
depend, however, on the renormalization point $\mu$ and on the
kinematic invariant $w=v_\c\cdot v_\b$. From here on we will suppress
these arguments.

Performing an integration by parts on the interaction lagrangian and
using the equations of motion (see ref.~[\mike]
for details), we find that
in matrix elements
$$ \ol h_\c\big[-\im v_\b\cdot\overleftarrow{\cal{D}}(v_\c)\big]
   \Gamma h_\b \to (\lambar - \dm)(w-1)\,\ol h_\c\Gamma h_\b ~.
   \eqn\lambarintro$$
Here $\lambar$ denotes the difference between the mass of the hadron
containing the heavy quark and $m_Q$, \ie\ in this case
$\lambar=m_B-m_\b=m_D-m_\c$. The relation \lambarintro\ implies that,
in matrix elements, we may replace
$$ Q_2 \to {1\over2m_\c}(\lambar-\dm)(w-1)\, \ol h_\c\Gamma h_\b ~.
   \eqn\qthreerel$$
The matrix elements of $Q_2$ (and $Q_4$, which obeys an analogous
relation) are thus proportional to the Isgur-Wise function $\xi_0$.
Furthermore, we see that they involve the combination $(\lambar-\dm)$,
which is invariant under shifts of $m_\c$ and $m_\b$.

As in ref.~[\mike], the equations of motion and relation \lambarintro\
may be used to reduce the number of independent form factors.  We find
$$ \eqalign{
   \xi_- &= {1\over2}(\lambar-\dm)\,\xi_0 ~,\cr
   (w+1)\,\xi_+ &= {1\over 2}(w-1)(\lambar-\dm)\,\xi_0 - \xi_3 ~.\cr}
   \eqn\xirels$$
The corrections proportional to $1/m_\b$, although we do not present them
explicitly, involve the same set of universal
functions since the second of
eqs.~\traceform\ can be related by an integration
by parts to a formula for
the matrix elements of $\ol h_\c\Gamma\im{\cal{D}}_\mu(v_\b) h_\b$.

The $1/m_\c$ corrections to the hadronic wave functions come from
insertions of the subleading operators
$$ {1\over2m_\c}\,\ol h_\c(\im{\cal{D}})^2h_\c
   + a_\c(\mu){g\over4m_\c}\,\ol h_\c\sigma^{\mu\nu}G_{\mu\nu} h_\c
   \equiv \CO_1+a_\c(\mu)\CO_2
   \eqn\odefs$$
into matrix elements of the leading order currents. This gives rise to
additional unknown functions.
In particular, insertions of $\CO_2$ induce
violations of the heavy quark spin symmetry.
Extending the trace formalism
to these contributions [\mike], we defined Lorentz and CP invariant
form factors $\chi_i$ via the time-ordered products\footnote\dag{This
definition of $\chi_2$ and $\chi_3$ differs from that of ref.~[\mike] by
the prefactor $a_\c(\mu)$.}
$$ \eqalign{
   \braa{D^{(*)}(v_\c)} &|\;\im\!\int\! d^4y\,
   T\big[\,\ol h_\c\Gamma h_\b(0),\CO_1(y)\,\big]\;|\kett{B^{(*)}(v_\b)}
   \cr
   &= -{\chi_1(w,\mu)\over m_\c}\,\Tr\big[\,\ol M_D\Gamma M_B\,\big]
   ~, \cr
   \braa{D^{(*)}(v_\c)} &|\;\im\!\int\! d^4y\,
   T\big[\,\ol h_\c\Gamma h_\b(0),\CO_2(y)\,\big]\;|\kett{B^{(*)}(v_\b)}
   \cr
   &= -{1\over m_\c}\,\Tr\Big[\big(\im\chi_2(w,\mu)\,\gamma^\mu v^\nu_\b
   + \chi_3(w,\mu)\,\sigma^{\mu\nu}\big)\,\ol M_D\sigma_{\mu\nu}
   {(1+\vslash_\c)\over2}\Gamma M_B\Big] ~.\cr}
   \eqn\chidefs$$
Hence to order $1/m_\c$ the hadronic matrix elements of the weak current
may be expressed in terms of five universal (real) functions of $w$ and
the parameter $(\lambar-\dm)$. As mentioned above, the corrections
proportional to $1/m_\b$ involve the same functions [\Volker].

Evaluating the traces in \traceform\ and \chidefs, we obtain the vector
and axial vector current matrix elements relevant for semileptonic
$B\to D^{(*)}$ decays. To order $1/m_\c$,
they are\footnote\ddag{We correct
an error in expressions (3.11) and (3.14) of ref.~[\mike].}
$$ \eqalign{
   \braa{D}|\,V^\mu\,|\kett{B}=\sqrt{m_D m_B}\,
   \Big\{ &c_0\,\xi_0\,(v_\c^\mu+v_\b^\mu) \cr
   + &{c_1\over2m_\c}\,\big[\xi_0\,(\lambar-\dm)-2\xi_3\big]\,
   (v_\c^\mu-v_\b^\mu) \cr
   + &{c_2\over2m_\c}\,\xi_0\,(\lambar-\dm)(w-1)\,
   (v_\c^\mu+v_\b^\mu) \cr
   + &{\c_0\over2m_\c}\,\big[ 2\chi_1-4(w-1)a_\c\,\chi_2+
   12a_\c\,\chi_3\big]\,(v_\c^\mu+v_\b^\mu)\Big\} ~,\cr}
   \eqn\vpscpsc$$
$$ \eqalign{
   \braa{D^*(\eps)}|\,V^\mu\,|\kett{B}=\im&\sqrt{m_{D^*} m_B}\,
   \eps^{\mu\nu\alpha\beta}\,\eps^*_\nu v_{\c\alpha} v_{\b\beta}\,
   \Big\{ c_0\,\xi_0 + {c_1\over2m_\c}\,\xi_0\,(\lambar-\dm) \cr
   & + {c_2\over2m_\c}\,\xi_0\,(\lambar-\dm)(w-1)
   + {c_0\over2m_\c}\,\big[ 2\chi_1-4a_\c\,\chi_3\big]
   \Big\} ~, \cr}
   \eqn\vpscvec$$
$$ \eqalign{\braa{D^*(\eps)}|\,A^\mu\,|\kett{B} =&\sqrt{m_{D^*} m_B}\,
   \Big\{c_0\,\xi_0\,\big[\eps^{*\mu} (w+1)-\eps^*\!\cdot\!v_\b\,
   v_\c^\mu\big] \cr
   &+ {c_1\over2m_\c}\,\Big[{2\over w+1}
   \big[\xi_0\,(\lambar-\dm)+\xi_3\big]\,\eps^*\!\cdot\! v_\b\,
   (v_\c^\mu+v_\b^\mu) \cr
   &\qquad\qquad + \xi_0\,(\lambar-\dm)\,\big[(w-1)\eps^{*\mu} -
   \eps^*\!\cdot\!v_\b\,v_\c^\mu\big] \Big] \cr
   &+ {c_2\over2m_\c}\,\xi_0\,(\lambar-\dm)(w-1)\,
   \big[\eps^{*\mu}(w+1)-\eps^*\!\cdot\!v_\b\,v_\c^\mu\big] \cr
   &+ {c_0\over2m_\c}\,\Big[ 4a_\c\,\chi_2\,\eps^*\!\cdot\!v_\b
   \,(v_\c^\mu-v_\b^\mu) \cr
   &\qquad\qquad +\big[ 2\chi_1-4a_\c\,\chi_3\big]
   \big[\eps^{*\mu}(w+1)-\eps^*\!\cdot\!v_\b\,v_\c^\mu)\big]\Big]\Big\}
   ~. \cr}
   \eqn\apscvec$$
It is clear that the requirement that these expressions be invariant
under redefinitions of the expansion parameters $m_\c$ and $m_\b$ is
equivalent to the requirement that the form factors be invariant, \ie
$$ {\partial\xi_i\over\partial m_Q} =
   {\partial\chi_i\over\partial m_Q} = 0 ~,
   \eqn\xicond$$
where $m_Q$ is any of the heavy quark masses.
This is not unexpected, since we have defined the form factors in
terms of matrix elements of operators built from the ``generalized''
covariant derivative ${\cal{D}}$. It is these form factors,
together with the invariant combination $(\lambar-\dm)$, which are
observables of the effective theory.

We recall that by evaluating matrix elements of the charge operator
$V_0$ at zero recoil one can show that the Isgur-Wise function is
normalized at zero recoil, $\xi_0(1,\mu)=1$, and that two of the
subleading form factors vanish at this point, $\chi_1(1,\mu)=
\chi_3(1,\mu)=0$, if one uses the hadron masses in the prefactors in
\vpscpsc--\apscvec. These conditions are preserved in the effective
theory with a residual mass term.

The physical matrix elements must be invariant not only under shifts
in the expansion parameter $m_Q$, but also under changes in the
renormalization point $\mu$.  We can use this fact to deduce the
$\mu$-dependence of the universal functions $\xi_0$, $\xi_3$ and
$\chi_i$ from that of the short distance coefficients $c_i$. To first
order in $\alpha_s$ we find that
$$ \eqalign{
   \mudmu{\xi_i} &=-{4\alpha_s\over3\pi}\,[wr(w)-1]\,\xi_i ~;~~
   i=0,3 ~, \cr
   \mudmu{\chi_1} &=-{4\alpha_s\over3\pi}\,\bigg\{ [wr(w)-1]\,\chi_1
   - \xi_0\,(\lambar-\dm)\,{r(w)-w\over w+1}\bigg\} ~, \cr
   \mudmu{\chi_i} &=-{4\alpha_s\over3\pi}\,\Big[wr(w)+\fr18\Big]\,
   \chi_i ~;~~ i=2,3 ~. \cr}
   \eqn\mudepend$$
These conditions must be obeyed by any model calculation which is
sensitive to the $\mu$-dependence, such as leading-log improved QCD sum
rules.

\section{4. Heavy-Light Currents}

The arguments given above apply equally to currents with one heavy and
one light quark.  We have verified that again the only change with
respect to the case $\dm=0$ consists in the replacement of the
covariant derivative $D_\mu$ acting on a heavy quark field by
${\cal{D}}_\mu$ from \deriv. Derivatives acting on the light quark
field remain unchanged. The study of heavy-light currents is
interesting, however, since it will allow us to give an operational
definition of the invariant combination $(\lambar-\dm)$ which does not
contain the residual mass explicitly.

In leading logarithmic approximation, the $1/m_Q$ expansion of the
current $J=\ol\q\Gamma Q$ in the effective theory reads
$$ J_{\rm eff}=b_0(\mu)\,Q'_0 + {1\over 2 m_Q}
   \sum_{i=1}^6 b_i(\mu)\,Q'_i ~,
   \eqn\heavylight$$
where
$$ Q'_0 = \e^{\im\phi'}\,\ol\q\,\Gamma h_Q ~,~~
   Q'_1 = \e^{\im\phi'}\,\ol\q\,\Gamma\,
   \im\,\rlap/\!{\cal{D}} h_Q ,
   \eqn\opdefs$$
and $\phi'=-m_Q v\cdot x$.  The remaining operators in \heavylight\
involve the light quark mass $m_\q$ or derivatives acting on the light
quark fields. They cannot easily be written in terms of an arbitrary
$\Gamma$.  For $\Gamma=\gamma_\mu$, it is convenient to choose
$$ \eqalign{
   Q'_2 &= \e^{\im\phi'}\,m_\q\,\ol\q\,\gamma_\mu h_Q ~, \cr
   Q'_3 &= \e^{\im\phi'}\,m_\q\,\ol\q\,v_\mu h_Q ~, \cr
   Q'_4 &= \e^{\im\phi'}\,\ol\q\,(-\im\overleftarrow{D}_\mu) h_Q ~,
   \cr}
   \qquad
   \eqalign{
   Q'_5 &= \e^{\im\phi'}\,\ol\q\,(-\im v\cdot\overleftarrow{D})
   \gamma_\mu h_Q ~, \cr
   Q'_6 &= \e^{\im\phi'}\,\ol\q\,(-\im v\cdot\overleftarrow{D})
   v_\mu h_Q ~. \cr}
   \eqn\moreops$$
We neglect again operators whose matrix elements vanish by the equations
of motion.  A set of axial vector operators with the same coefficients
$b_i(\mu)$ is obtained from \opdefs\ and \moreops\ by replacing
$\ol\q\to-\ol\q\gamma^5$ and $m_\q\to-m_\q$. From tree level matching
one computes $b_0=b_1=1$ and $b_i=0$ for $i\ge2$, while at leading
logarithmic order (this may be extracted from ref.~[\FG])
$$ \eqalign{
    b_0(\mu) &= b_1(\mu) = z^{-6/\beta} ~, \cr
    b_2(\mu) &= -\fr{10}9-\fr19z^{6/\beta}+\fr89z^{3/\beta}
     +\fr13z^{-6/\beta} ~, \cr
    b_3(\mu) &= \fr{10}9-\fr{26}9z^{6/\beta}+\fr49z^{3/\beta}
     +\fr43z^{-6/\beta} ~, \cr
    b_4(\mu) &= \fr{10}3-\fr43z^{3/\beta}-2z^{-6/\beta} ~, \cr
    b_5(\mu) &= -\fr{10}9-\fr4{27}z^{3/\beta}+\fr{34}{27}z^{-6/\beta}
     -\fr{16}\beta z^{-6/\beta}\ln z ~, \cr
    b_6(\mu) &= -\fr{20}9+\fr{88}{27}z^{3/\beta}
     -\fr{28}{27}z^{-6/\beta} ~, \cr}
   \eqn\bdefins$$
where $z=\alpha_s(m_Q)/\alpha_s(\mu)$ and $\beta=33-2N_f$.

Let us now consider matrix elements of the effective current between a
heavy meson and the vacuum. The application of the trace formalism to
this particular case allows us to write
$$ \eqalign{
   \braa{0}|\,\ol\q\,\Gamma h_Q\,|\kett{M(v)} &= \im\,{F(\mu)\over 2}\,
   \Tr\big[\,\Gamma\,M\,\big] ~, \cr
   \braa{0}|\,\ol\q\,\Gamma\,\im{\cal{D}}_\mu\,h_Q\,|\kett{M(v)}
   &= {\im\over 2}\,\Tr\big[\,\big(F_1(\mu) v_\mu + F_2(\mu)
   \gamma_\mu\big)\,\Gamma\, M\,\big] ~, \cr
   \braa{0}|\,\ol\q\,(-\im\overleftarrow{D}_\mu)\Gamma h_Q\,|\kett{M(v)}
   &= {\im\over 2}\,\Tr\big[\,\big(F_3(\mu) v_\mu + F_4(\mu)
   \gamma_\mu\big)\,\Gamma\,M\,\big] ~. \cr}
   \eqn\Fdefs $$
Using an integration by parts and the
equations of motion of the effective
theory, one can show that [\matthias]
$$ \eqalign{
   F_1(\mu) &= F_2(\mu) = F_4(\mu)
   = -{F(\mu)\over 3}\,\big[\lambar-\dm-m_\q\big] ~. \cr
   F_3(\mu) &= -{F(\mu)\over 3}\,\big[ 4(\lambar-\dm)-m_\q\big] ~. \cr}
   \eqn\Fres $$
Additional parameters $G_1(\mu)$ and $G_2(\mu)$ are induced by
insertions of subleading operators from the effective lagrangian
\lkineff, in analogy to the functions $\chi_i(w,\mu)$.

It is clear from \Fres\ that physical matrix elements depend only on
the combination $(\lambar-\dm)$ and the form factors $F, G_1$, and
$G_2$, which must therefore be invariant under redefinitions of $m_Q$.
As an example, we present the result for the ratio of the decay
constants of a heavy vector and pseudoscalar meson [\matthias]
$$ \eqalign{
   {f_{M^*}\sqrt{m_{M^*}}\over f_M\sqrt{m_M}}
   =1 &+ {1\over2m_Q}\,\Big\{ (\lambar-\dm-m_\q)\,c_0^{-1}(\mu)\,
   \big[\fr43b_1(\mu)+\fr23b_4(\mu)+b_6(\mu)\big] \cr
   &+ m_\q\,c_0^{-1}(\mu)\,\big[2b_2(\mu)+b_3(\mu)+b_4(\mu)+b_6(\mu)\big]
   -16G_2(\mu)\Big\} ~. \cr}
   \eqn\ratio$$
 From the fact that this ratio is $\mu$-independent we conclude that the
scale dependence of $G_2$ involves $m_\q$ and $(\lambar-\dm)$, in analogy
to \mudepend.

The relations \Fres\ allow us to give an operational definition of the
combination $(\lambar-\dm)$, which is invariant under redefinitions of
$m_Q$, without reference to the residual mass term. The reason is that
this combination is induced when the {\it usual\/} covariant derivative
acts on the light quark. In particular, it follows that
$$ {\braa{0}|\,\ol\q\,(\im v\cdot\overleftarrow{D})\Gamma h_Q\,|
    \kett{M(v)} \over
    \braa{0}|\,\ol\q\,\Gamma h_Q\,|\kett{M(v)}}
   = \lambar-\dm ~.
   \eqn\definition $$
Note that this relation expresses $(\lambar-\dm)$ directly in terms of
properties of the light degrees of freedom, instead of in terms of a
difference of heavy masses. We may even propose \definition\ as a
definition of the energy of the light degrees of freedom in the
background of a static color source. This equation might also be useful
in extracting the value of $(\lambar-\dm)$ from lattice gauge theory.

\section{5. Summary}

We have demonstrated in detail that there is no important
ambiguity in the choice of the HQET expansion parameter $1/m_Q$, in the
sense that all physical quantities are independent of this choice. This
should hardly be surprising.  The ambiguity is resolved by the
introduction of a new quantity $\dm$, which is a non-trivial dynamical
parameter of the effective theory. In order to be insensitive to the
choice of the expansion parameter $m_Q$, hadronic form factors in the
HQET must be defined in terms of matrix elements containing the operator
$\im{\cal{D}}_\mu=\im D_\mu-\dm v_\mu$. Physical quantities will then
depend on these invariant form factors as well as on the combination
$(\lambar-\dm)$, which replaces $\lambar$ in previous analyses. This
combination is invariant under redefinitions of $m_Q$.

While calculable in continuum perturbation theory, the residual mass
term contains power divergences when the theory is regulated with a
dimensionful cutoff, divergences which may be associated with
incalculable nonperturbative processes [\mms]. The issue of whether such
effects make the HQET useless beyond leading order in $1/m_Q$ is beyond
the scope of this letter. It is important, however, to disentangle this
larger question from that of whether the theory is plagued by a
fundamental ambiguity in the choice of expansion parameter.  We have shown
that it is not, in the presence of the residual mass $\dm$. We believe
that this is the correct framework within which to investigate further
the possible importance of nonperturbative effects.

\bigskip
\ACK It is a pleasure to thank Nathan Isgur, Chris Sachrajda and Guido
Martinelli for provocative questions and interesting discussions.
Portions of this work were initiated at the 1992 Santa Barbara Workshop
on Heavy Quark Symmetry, and we are most grateful to the Institute for
Theoretical Physics for their kind hospitality.
One of us (M.N.) gratefully
acknowledges financial support from the Research Fellowship Program of
the BASF Aktiengesellschaft and the German
National Scholarship Foundation.

\refout

\bye